\documentstyle[aps,epsfig]{revtex}
\begin{document} 
\draft
\title{Event by Event Analysis and Entropy of Multiparticle Systems }

\author{A.Bialas and W.Czyz}
\address{%
M.Smoluchowski Institute of Physics, Jagellonian
University, Cracow
\\
Reymonta 4, 30-059 Krakow, Poland;
e-mail:bialas@thp1.if.uj.edu.pl
\\
and
\\
H.Niewodniczanski Institute of Nuclear Physics
\\
Radzikowskiego 152, Krakow, Poland}
\maketitle

\begin{abstract} 
The coincidence method of measuring the entropy of a system, proposed
some time ago by Ma, is generalized to include systems out of
equilibrium. It is suggested that the method can be adapted to analyze multiparticle states 
produced in high-energy collisions.
\end{abstract}
 
{\bf 1.} Entropy, being one of the most important characteristics of a
system with many degrees of freedom, is --- in particular --- an important
characteristics of multiparticle production processes. In this
context it abounds in analyses of dense hadronic matter and  in discussions of
various models of quark-gluon plasma \cite{qm}.

Processes in which particles are produced can be considered as the so
called {\it dynamical systems} \cite{gu,mi} in which 
--- generally --- entropy
gets produced. Although application of the mathematical theory of
dynamical systems to calculate the entropy in multiparticle production
is still out of reach, the existing models suggest that the systems
produced in high-energy collisions pass through a stage of (approximate)
local statistical equilibrium \cite{eq}.

Recently \cite{bcw} we have proposed to apply the event coincidence method
\cite{ma} to measure entropy of a multiparticle systems, provided it can
be described by a microcanonical ensemble\footnote{A direct measurement
of entropy of multiplicity distribution observed in multiparticle
production was first reported in \cite{si}.}.
 Since the event-by-event analysis becomes a commonly 
accepted tool to study the multiparticle phenomena, we feel that it is
worth to pursue this problem further. 
 In the present paper we
extend the coincidence method to the more realistic case when the energy of the
system in question is not necessarily fixed. We show that the method can
be rather effective  investigating  {\it local} properties of the
particle spectra. Since the observed particles map the state of the
system just before it breaks into freely-moving hadrons (which get
registered in the detectors), such a measurement 
 can provide an important information on the
evolution of the system\footnote {Note that the free movement of
particle from production point to the detector does not influence this
measurement.}.

At this point it may be important to stress that to estimate properly
the entropy of a multiparticle system one would need information not
only on distribution of momenta but also about positions of particles.
In particular, correlations between positions and momenta are very
essential. This information cannot be obtained, generally, in a
model-independent way. One should thus keep in mind that the entropy we
discuss in the present paper reflects only partially the statistical
properties of the system: the degrees of freedom related to positions of
particles are integrated over. Nevertheless it provides a valuable
information about the system in question, and can be used to identify
its nature. In particular, our method may have a wide range of
application for the systems where correlations between momenta and
positions of the particles are unimportant.

{\bf 2.} In a system at equilibrium with all states having the same
probability (microcanonical ensemble) entropy measures the number $\Gamma$ of
states of the system: 
\begin{equation}
 S= \log\Gamma. \label{1}
 \end{equation} 
This formula can be rewritten in terms
of the probability $p$ for one of the states of the system to realize.
Since all states have equal probabilities we have 
\begin{equation}
p=\frac1{\Gamma}\label{2}
 \end{equation}
 and thus 
\begin{equation}
 S= -\log p. \label{3} 
\end{equation} 
Ma observed \cite{ma} that the
probability $p$ can also be expressed as probability of "coincidence",
i.e. probability that while sampling the system, one finds two states
(configurations) which are identical to each other. Indeed, this
probability is given by 
\begin{equation} 
C_2 = \sum _{\rm all\; states}
(p^2) = \Gamma p^2 = p \label{4}
 \end{equation}
 so that 
\begin{equation}
S=-\log C_2. \label{4a}
 \end{equation}
 Now, if we measure $N$
configurations and find $N_2$ coincidences we have (in the limit of
large $N$)
 \begin{equation} 
C_2 = \frac {N_2}{N(N-1)/2} \label{5}
\end{equation}
 and thus we obtain a method of estimating $p$ and
therefore also of entropy $S$. The attractive feature of this procedure
is that, as seen from (\ref{5}), the statistical error drops very fast
(like $N^{-1})$ with increasing number of the tried configurations\footnote{
This holds for $N$ in the region $\sqrt{\Gamma} \ll N \ll
\Gamma$, the case of interest in the present context.}.

This method does not work, however, if the energy of the considered
system is not precisely fixed (e.g. for canonical or grand-canonical
ensemble) or if the system is not in termodynamic equilibrium. In such
a case the states of the system have, in general, various probabilities
of occurence. Consequently, neither (\ref{3}) nor  (\ref{4}) are valid. 

In the present note we  argue that even in this general case 
the coincidence
method can nevertheless be used to obtain information on the entropy of
the system. To this end it is, however, necessary to measure concidences 
of more than two configurations. The argument goes as follows.

For an arbitrary system  entropy
is  defined by the general formula \cite{sh}
\begin{equation}
S=-\sum_{{\rm all}\;n} p_n \log p_n,   \label{6}
\end{equation}
where $p_n$ is the probability of occurence of the state labelled by
$n$, and the sum runs over all states of the system.

To begin we observe that (\ref{6}) can be rewritten as
\begin{equation}
S=-\left< \log p \right>,   \label{7}
\end{equation}
where $<...>$ denotes the average over all states of the system.

Using now the identity
\begin{equation}
p =<p>\frac{p}{<p>} = <p> \left[1-\left(1- \frac{p}{<p>}\right)\right]
  \label{11}
\end{equation}
one can transform (\ref{7}) into
\begin{equation}
S= - \log<p> + \sum_{m=2}^{\infty}\frac{1}{m} \left<
 \left(1- \frac{p}{<p>}\right)^m\right>. \label{12}
\end{equation}
In this way we have expressed the entropy by the moments 
$<p^m>$. 

Now, the point is that these moments have a simple physical
interpretation in terms of the coincidence probability. Indeed, 
let us denote by $C_k$ the probability of coincidence of $k$
configurations. In terms of probabilities $p_n$ it can be  expressed as
\footnote{This formula can be easily proven by considering the Bernoulli 
distribution of $N$ independent samplings of the considered system. The error
can be estimated with the same technique.}:
\begin{equation}
C_k = \sum _{{\rm all}\; n} \left( p_n\right) ^k 
=\sum _{{\rm all}\; n} p_n \left( p_n\right) ^{k-1} = <p^{k-1}>.
 \label{8}
\end{equation}

 We see that  the probability
of  coincidence of $k$ configurations is given by 
 the $k-1$-th moment of $p$.

We thus conclude from (\ref{12})  that the
 probabilities $C_k$ of coincidences of all orders   
are in principle necessary to 
determine the entropy of the system. 

In terms of $C_k's$, (\ref{12}) reads
\begin{equation}
S= - \log C_2 + \sum_{m=2}^{\infty}\frac{1}{m} 
\sum_{k=0}^{m}(-1)^k {m\choose k}\frac{C_{k+1}}{(C_2)^k}.
 \label{13}
\end{equation}
If all states have the same probability of occurence we obtain trivially
$C_{k+1}= (C_2)^k$. Thus all terms in the sum vanish and we fall back to
the formula (\ref{4a})\footnote{$\sum_{k=0}^m(-1)^k{m\choose k}=(1-1)^m=0$.}.

Of course the series (\ref{13}) and its approximations may be used for
estimation of entropy only if 
the result is convergent. To this end the consecutive terms must
be small enough and thus the parameters $C_{k+1}/(C_2)^k$ cannot be much
larger than one\footnote{It is not difficult to see that
$C_{k+1}/(C_2)^k \geq 1.$ Indeed, for any positive variable $f$ we
have $<f^{k-1}(f-<f>)^2> \geq 0.$. It follows that $<f^{k+1}>-<f>^{k+1}
\geq 3 <f>^2 (<f^{k-1}>-<f>^{k-1})$ and one can complete the proof by
induction.}. This condition limits seriously the applicability of (\ref{13}).

{\bf 3.} It is  useful to rearrange the  series  (\ref{13})
 using the  so-called
replica method \cite{pa}. To
this end, let us consider a system made of M independent replicas of
the considered system. The entropy of such a composite system is obviously
given by
\begin{equation}
S(M) = M S.   \label{2.1}
\end{equation}
On the other hand, since it is made of $M$ independent subsystems the
coincidence probabilities are given by
\begin{equation}
C_k(M) = [C_k]^M.    \label{2.2}
\end{equation}
Consequently,  repeating the argument of the previous section we obtain
\begin{equation}
S(M)= -M \log C_2 + \sum_{m=2}^{\infty}\frac{1}{m} 
\sum_{k=0}^{m}(-1)^k {m\choose k}\left(\frac{C_{k+1}}{(C_2)^k}
\right)^M.
 \label{2.3}
\end{equation}
Now, consistency of (\ref{2.1}) and (\ref{2.3}) requires that the sum on
the R.H.S. of (\ref{2.3}) is proportional to $M$ and thus only the term
proportional to $M$ can survive. This term is easy to calculate by
observing that
\begin{equation}
\left(\frac{C_{k+1}}{(C_2)^k}\right)^M = 1+ M \log\left(\frac{C_{k+1}}{(C_2)^k}
\right) +...\,.    \label{2.4}
\end{equation}
Substituting this into (\ref{2.3}) we
obtain
\begin{eqnarray}
S(M)= -M \log C_2 + M\sum_{m=2}^{\infty}\frac{1}{m} 
\sum_{k=0}^{m}(-1)^k {m\choose k}\log
\left(\frac{C_{k+1}}{[C_2]^k}\right).
 \label{2.5}
\end{eqnarray}

Using (\ref{2.1}) we thus have
\begin{equation}
S= -\log C_2 + \sum_{m=2}^{\infty}\frac{1}{m} 
\sum_{k=2}^{m}(-1)^k {m\choose k}\log
\left(\frac{C_{k+1}}{[C_2]^k}\right)
 \label{2.7}
\end{equation}
which represents our final formula. It is
providing partial resummation of the powers of $C_{k+1}/[C_2]^k$ into
logarithms.

{\bf 4.} The formula (\ref{2.7}) can be rewritten in terms of the Renyi
entropies\footnote{The argument presented in this section 
 was suggested to us by K.Zyczkowski.} defined as \cite{re}
\begin{equation}
H_k= -\frac{\log C_k}{k-1}.  \label{7.1a}
\end{equation}
Using this definition one can easily see that $H_1= S$. 
Substituting (\ref{7.1a}) into (\ref{2.7}) we obtain after some algebra
\begin{eqnarray}
S= H_2 + \sum_{n=1}^{\infty}\sum_{k=0}^n (-1)^k{n\choose k}H_{k+2}
=\sum_{n=0}^{\infty}\sum_{k=0}^n (-1)^k{n\choose k}H_{k+2} \nonumber\\
= H_2 +(H_2-H_3) +(H_2-3H_3+H_4)+(H_2-3H_3+3H_4-H_5)+...\,.  \label{7.2a}
\end{eqnarray}
One sees that the first $N$ terms of this series represent 
the polynomial extrapolation
 of the function $H_k$ from the points $k=2,3,4,..,N+1 $ to $k=1$. This
observation not  only explains the meaning of  formulae (\ref{2.7}) 
and (\ref{7.2a}) but also
suggests the way to improve it: one should look for more effective
extrapolations. One  possibility we have investigated in some detail 
is to take
\begin{equation}
H_k=  a \frac{\log k}{k-1} + a_0+a_1(k-1) + a_2(k-1)^2 +...\,.  \label{7.3a}
\end{equation}
Number of terms is determined by the number of coincidence probabilities
one is able to measure. If only $C_2$ and $C_3$ are measured we obtain
\begin{equation}
S=H_2 +\frac{1-\log 2}{\log 2 -(1/2) \log3}(H_2-H_3). \label{7.4a}
\end{equation}
If three coincidences are measured we have
\begin{equation}
S=H_2 +(H_2-H_3)(1+\omega) - \omega(H_3-H_4),  \label{7.5a}
\end{equation}
where
\begin{equation}
 \omega= \frac{1-2\log 2 + (1/2)\log 3}{\log (2/3) +(2/3)\log 2}. \label{7.6a}
\end{equation}
In Figure 1 the results of this procedure are shown for three
distributions, often encountered in the analysis of multiparticle data: 
Poisson, Negative Binomial and the Geometric series. One sees that
extrapolation using only two terms is by far sufficient to obtain an accurate
value of entropy, provided the average multiplicity is not lower than
$1/2$. The first term ($H_2$) is, however, hardly sufficient even for fairly
large multiplicities.

For $\bar{n} \rightarrow 0$ the extrapolation is rather poor which shows
that the method is not well adapted for studies of low multiplicity
events.
 
We have also found that for these three distributions the polynomial
extrapolation (\ref{7.2a}) less accurate than (\ref{7.3a}).

{\bf 5.} We have suggested recently \cite{bcw} that the coincidence
method of Ma can be used to estimate the entropy of the system of
particles produced in a high-energy collision. The idea was to consider
the produced events as the randomly chosen configurations of the system.
Measurement of the (appropriately defined) probability of coincidence of
two events was interpreted, following the formula (\ref{4a}), as a
measurement of entropy of the system\footnote{As
explained in Section 1, we are considering only entropy related
to the distribution of particle momenta. The volume fluctuations and
correlations between the position and momentum of a particle are neglected.}.

As it is not very likely that the system produced in a high-energy
collision can be indeed accurately represented by a microcanonical
ensemble at equilibrium \footnote{Although this is the case in the Fermi
model of multiparticle production \cite{f}.}, however, one may have
justified doubts about the accuracy of this method. It is clear from the
previous argument that the Eqs.(\ref{2.7}) and (\ref{7.2a}) provide a
possibility to assess this. Indeed, already measuring the probability of
coincidence of three events
\begin{equation}
C_3 = \frac{N_3}{N(N-1)(N-2)/6}      \label{15}
\end{equation}
allows one to estimate the first correction to the Eq.(\ref{4a}). As
discussed in the previous section, this is often sufficient to obtain
an accurate value of the entropy.

{\bf 6.} Application of the coincidence method, as described in 
previous sections, for measurements of entropy in multiparticle
production (which is our main objective) requires discretization
of the observed multiparticle spectra \cite{bcw}. The 
dependence of the results of measurements on discretization
 can be discussed as follows.

Consider a system consisting of a certain number, say $N$, of particles
produced in a high-energy collision. Let $\Phi(q)dq\equiv
\Phi(q_1,..q_N)dq_1...dq_N$ be their probability distribution in
momentum space. To discretize, we split the distribution into M (3N
dimensional) bins of size $\Delta q_m$ , $m=1,...,M$. The
probability distribution to find the system in the bin $m$ is
\begin{equation}
w(m,M)= \Phi(q^{(1)}(m),...,q^{(N)}(m))\Delta q_{m},   \label{5.1}
\end{equation}
where $[q^{(1)}(m),...,q^{(N)}(m))]$ is the set of $N$  momenta
 defining the bin $m$. The coincidence probabilities measured from the
distribution (\ref{5.1}) are 
\begin{equation}
C_k(M)= \sum_{m=1}^M(\Delta q_m)^k [\Phi(q^{(i)}(m))]^k.   \label{5.2}
\end{equation}
If we now split each bin into $\lambda$ new bins (and thus multiply the
number of bins by factor $\lambda$) the probability (\ref{5.1}) changes accordingly 
and we obtain
\begin{equation}
C_k(\lambda M)=\frac1{\lambda^{k-1}} \sum_{m=1}^M(\Delta {q_m})^{k}
 \sum_{l_m=1}^{\lambda}
\frac1{\lambda} \left[\Phi(q^{(1)}(m,l_m),...,q^{(N)}(m,l_m))\right]^k.
 \label{5.3}
\end{equation}
For non-singular distribution $\Phi(q_1,...,q_N)$ the
 dependence of the sum on the R.H.S on $\lambda$
disappears in  the limit $\lambda \rightarrow \infty$ 
 and thus using  (\ref{2.7}) or (\ref{7.2a}) we have
\begin{equation}
S(\lambda M)=\log \lambda +S(M) \label{5.4}
\end{equation}
which summarizes the dependence of the proposed measurement on the
resolution used in the procedure of discretization\footnote{
Additional dependence on $\lambda$ would indicate that 
 the distribution $\Phi(q_1,...,q_N)$ is singular (see,e.g.,
 \cite{bp}).}. 
Note that $\lambda$ denotes the number of splittings in 3N dimensional
momentum space. If the splitting procedure is performed by simply
splitting the bins in one-dimensional single particle momentum distribution
into $\lambda_0$ new bins, we have
$\lambda = (\lambda_0)^{3N}$ which gives
\begin{equation}
S(\lambda M)=3N \log \lambda_0 +S(M). \label{5.4a}
\end{equation}

The final question one may ask is how the entropy measured from the
distribution (\ref{5.1}) is related to the "true" entropy\footnote{ We
use quotation mark to emphasize that, as explained in Section 1, the
entropy we discuss in this paper is not -in general-  the actual entropy of the system since
it neglects the positions of particles in configuration space.} of the
$N$ particle system described by the distribution function
$\Phi(q_1,...,q_N)$. To consider this problem we observe that the
spacing between the momentum states of a system of $N$ particles is
given by the quantum-mechanical relation
\begin{equation}
\delta q = \left(\frac{(2\pi)^3}{ v}\right)^N,   \label{5.5}
\end{equation}
where $v$ denotes the volume of the system\footnote{ The fluctuations 
of the volume 
can be -at least in principle- determined  if the HBT
correlations are measured for each event.}. Denoting the total number of
states of the system by $\Gamma$ the ``true'' entropy is given by
\begin{eqnarray}
S(\Gamma) = -\sum_{i=1}^{\Gamma} p(q^{(1)}(i),...,q^{(N)}(i)) \log
[p(q^{(1)}(i),...,q^{(N)}(i))]  \nonumber \\
= -\sum_{m=1}^M w(q^{(1)}(m),...,q^{(N)}(m))
\log [w(q^{(1)}(m),...,q^{(N)}(m))/\Gamma(m)]\nonumber \\ =S(M) +
\sum_{m=1}^M  w(q^{(1)}(m),...,q^{(N)}(m)) \log \Gamma(m), \label{5.6}
\end{eqnarray}
where
\begin{equation}
\Gamma(m)= \frac{  \Delta q_m}{\delta q}  = \left[\frac{(\Delta_0(M))^3}
{(2\pi)^3} v\right]^N \label{5.7}
\end{equation}
is the number of states in the bin $m$.
Here $\Delta_0(M)$ denotes the size of the (1-dimensional) bin in
momentum space of one particle.

Eq. (\ref{5.6}) relates the entropy $S(\Gamma)$ of the considered 
system to $S(M)$ - the one measured by discretization  into $M$ bins.
For
the simplest case when all bins used in discretization are equal to each
other, $\Gamma_m$ does not depend on $m$ and the last sum in (\ref{5.6})
can be performed. The result is
\begin{equation}
S(\Gamma) = S(M)+ \log(\Gamma(m))=
S(M) + 3N \log \left(v^{1/3} \frac{\Delta_0(M)  }{2\pi} \right).\label{5.8}
\end{equation}

{\bf 7.} To assess the practical possibilities of using the proposed
method to the actual multiparticle data, we have estimated the
coincidence probabilities for a system of particles produced
independently.

 Suppose that the produced particles come in a number of species,
labelled by $f$. Then
\begin{equation}
C_k = \prod_f C_k(f),   \label {7.1}
\end{equation}
so that it is enough to consider one kind of particles.

We now discretize the system by splitting it into $M$ bins of size $\Delta
q$. With this procedure, the state of the system is defined by giving the
number of particles in each bin. If particles are emitted intependently,
the probability of a given state is
\begin{equation}
W(n_1,....,n_M) = \prod_{i=1}^M P(n_i,\bar{n}_i),  \label{7.2}
\end{equation}
where $P(n,\bar{n})$ is the Poisson distribution with average $\bar{n}$
and $\bar{n}_i$ is the average number of particles in a bin labelled by
$i$ given by
\begin{equation}
\bar{n}_i = \int_{q_i-\Delta q/2}^{q_i+\Delta q/2} dq \rho(q),
\label{7.3}
\end{equation}
where $\rho(q)$ is the single particle momentum distribution:$ \int dq
\rho(q) = \bar {N }$ with $N$ being the total number of particles.

From (\ref{5.2}) we deduce 
\begin{equation}
C_k= \sum_{n_1,...,n_M} [W(n_1,...,n_M)]^k = \prod_{i=1}^M
C_k^{pois}(\bar{n}_i),    \label{7.4}
\end{equation}
where
\begin{equation}
C_k^{\rm pois}(\bar{n})=\sum_n [P(n,\bar{n})]^k.    \label{7.5}
\end{equation}

We have calculated numerically $C_k^{\rm pois}(\bar{n})$ for $2\leq k\leq
5$. They are shown in Fig.2, plotted versus $\bar{n}$. One sees that in
the range $1 \leq \bar{n} \leq 50$ they can be well approximated by the
formula 
\begin{equation}
C_k^{\rm pois}(\bar{n})\approx \left( \frac1{3\sqrt{\bar{n}}}\right)^{k-1}   
 \label{7.6}
\end{equation}
which shows that they are not prohibitively small even at fairly high
multiplicities. We thus conclude that {\it for one bin} at least $C_2$
and $C_3$ should be possible to measure with a reasonable accuracy even
for  large systems (i.e. systems containing many
particles\footnote{For large multiplicities the first term in the
asymptotic expansion of $C_k$ is $1/(\sqrt{2k\pi \bar{n}})^{k-1}$.}).

The situation becomes much worse, however, with the increasing number of
bins, as easily seen from (\ref{7.4}). For $\bar{N}=100$ and $M=10$
bins, for example, one obtains $C_2 \approx 10^{-9.5}$ and $C_3 \approx
10^{-19}$. The situation improves somewhat for smaller multiplicities:
$\bar{N}=10$ and $M=10$ one has $C_2 \approx 10^{-5}$ and $C_3 \approx
10^{-10}$. As shown in Section 6, however, the method does not work if
the particle multiplicity in one bin falls below $ \bar{n} \sim 1/2$.
Therefore it is limited to study of rather small regions of
phase-space.

{\bf 8.} At this point it may be worth to point out that the measurement
of event coincidence probabilities represents an interesting information
about the multiparticle system, independently of its relation to the
Shannon entropy. Indeed, it gives a valuable information on statistical
fluctuations of the system in question and thus may be considered as
alternative approach to the problem of "erraticity" \cite{hwa}. It seems
to be a more detailed measure of even-by-event fluctuations than the
distribution of the (horizontally averaged) factorial moments
\cite{hwa}. The weak point is that the method seems applicable only to a
small part of the available phase-space\footnote{An interesting
possibility would be to study two (or more) disconnected regions of
available phase-space, such a measurement being sensitive to the {\it
long range} correlations in the multiparticle system.}. Some averaging
procedure may thus turn out necessary also in this case. 

It is also worth to emphasize that the event coincidence probabilities
are sensitive to entirely different region of multiparticle spectrum
than the widely used factorial moments \cite{bp}. Indeed, whereas
factorial moments are sensitive mostly to the large multiplicity tail of
the spectrum, the coincidence probabilities obtain largest contributions
from the region where the probability distribution is maximal. The two
methods seem thus complementary to each other and should best be used in
parallel to obtain maximum of information.

{\bf 9.} In conclusion, we have proposed a generalization of Ma's
coincidence method of entropy determination. It requires measurements of
coincidences of $2,3,...$ configurations. The new method can be applied
to a more general class of systems. In particular, thermodynamical
equilibrium is not necessary. 

The method seems well adapted to analysis of local properties of
multiparticle states produced in high-energy collisions. It may thus
turn out useful for investigation of the thermodynamic properties of the
dense hadronic matter and/or quark-gluon plasma.

\vspace{0.3cm}
{\bf Acknowledgements}
\vspace{0.3cm}

We are greatly indebted to Karol Zyczkowski for the remarks which were
crucial for completing this paper. Discussions with Hans Feldmeier,
Hendrik van Hees, Jorn Knoll, Jacek Wosiek and Kacper Zalewski are
highly appreciated. AB thanks W. Noerenberg for the kind hospitality at
the GSI Teory Group where part of this work was done. This investigation
was supported in part by the KBN Grant No 2 P03B 086 14 and by Subsydium
of FNP 1/99.

\newpage

\begin{figure}
\epsfig{file=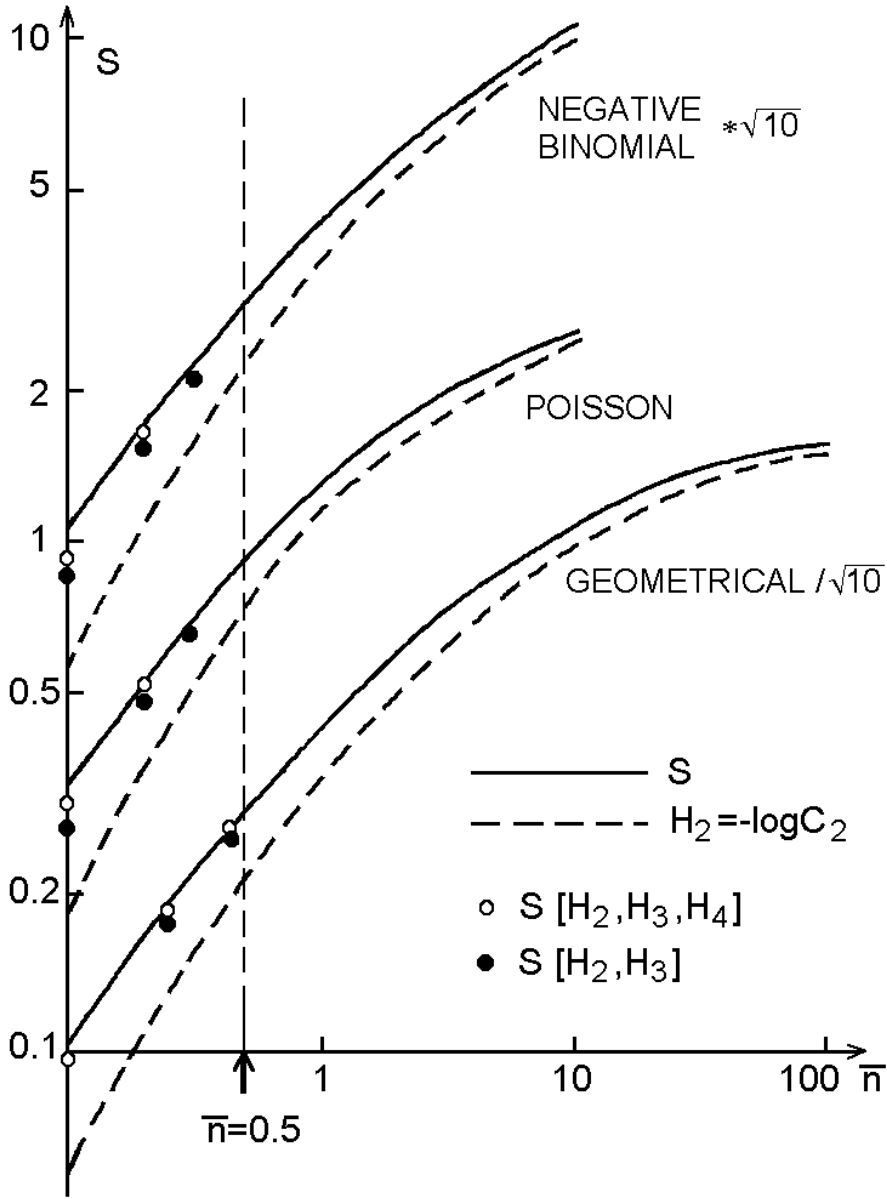,width=12cm}
\caption{Estimates of entropy for systems with commonly encountered
distributions using the extrapolation given by Eq.(\ref{7.3a}), plotted
versus average multiplicity. Continuous lines: entropy calculated
directly from (\ref{6}). Dashed lines: entropy calculated from
(\ref{4a}). Open points: Three-term extrapolation (\ref{7.5a}). Full points:
Two-term extrapolation (\ref{7.4a}).}
\end{figure}

\newpage

\begin{figure}
\epsfig{file=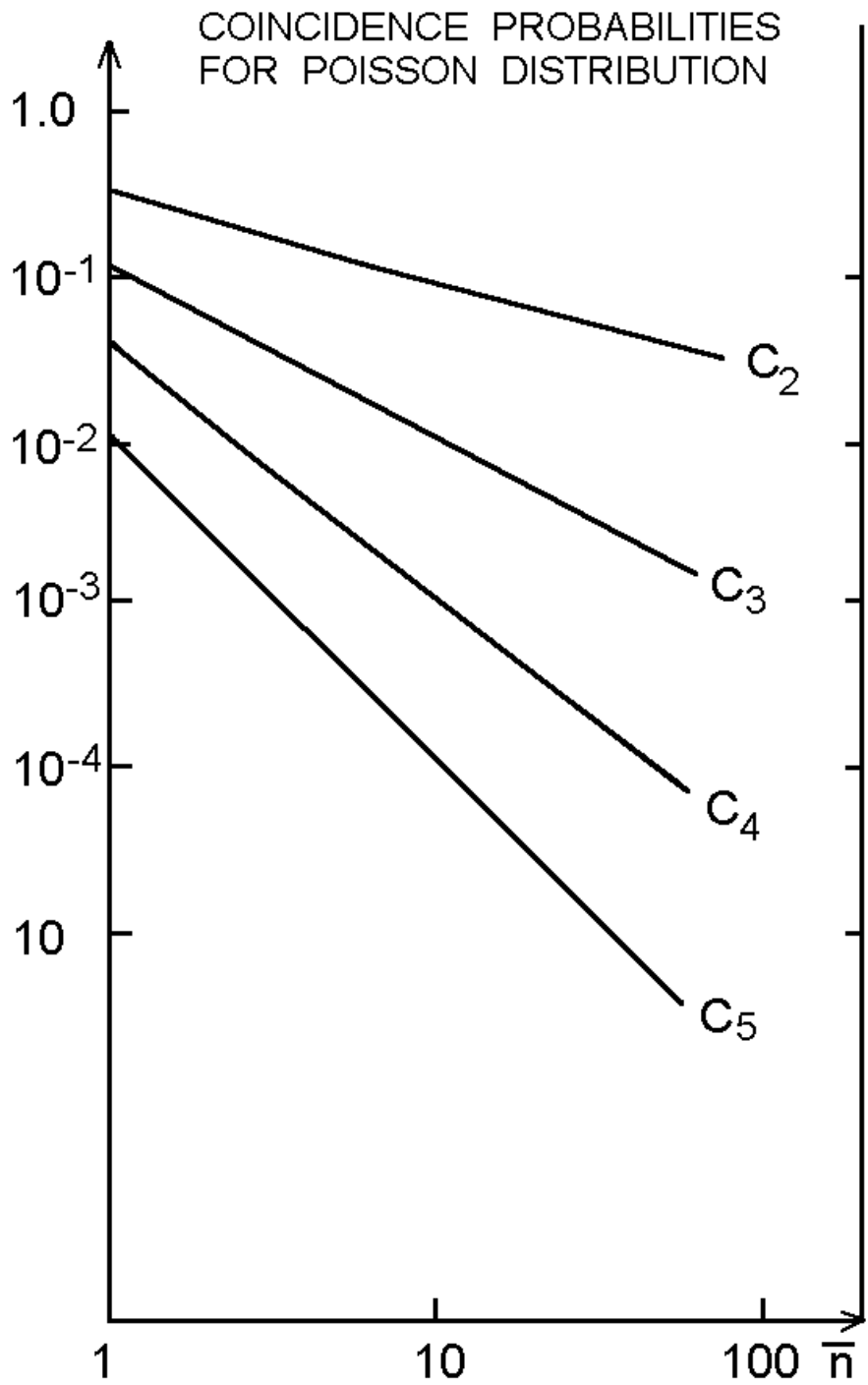,width=12cm}
\caption{Coincidence probabilities for Poisson distribution versus
average multiplicity.}
\end{figure}

\end{document}